\documentclass{elsart}

\usepackage[dvipdfm]{graphicx}

\usepackage{amssymb}

\begin{document}

\begin{frontmatter}



\title{
Quantum dynamics of N-methylacetamide studied by 
the vibrational configuration interaction method
}

\author[BU,GU]{Hiroshi Fujisaki\corauthref{pa}},
\author[TU,JST]{Kiyoshi Yagi},
\author[TU,JST]{Kimihiko Hirao}, and
\author[BU,MU]{John E. Straub}
\address[BU]{Department of Chemistry,
Boston University, 590 Commonwealth Avenue,
Boston, Massachusetts 02215, USA}
\address[GU]{Institute of Physical and Theoretical Chemistry,                                
J.W. Goethe University, Max-von-Laue-Str. 7,                                    
60438 Frankfurt am Main, Germany}   
\address[TU]{Department of Applied Chemistry, 
School of Engineering, The University of Tokyo, 
Hongo 7-3-1, Bunkyo-ku, Tokyo 113-8656, Japan}
\address[JST]{CREST, Japan Science and Technology Agency,
Saitama 332-0012, Japan}
\address[MU]{                                                                   
Department of Chemistry and Biochemistry,                                       
Montana State University, Bozeman, Montana 59717, USA                           
}                                                                               
\corauth[pa]{Corresponding author: fujisaki@theochem.uni-frankfurt.de}
              
\begin{abstract}


Vibrational energy transfer of the amide I mode of N-methylacetamide 
(NMA) is studied theoretically using the vibrational configuration 
interaction method. A quartic force field of NMA is constructed 
at the B3LYP/6-31G+(d) level of theory and its accuarcy is checked 
by comparing the resulting anharmonic frequencies with 
available theoretical and experimental values. Quantum dynamics 
calculations for the amide I mode excitation clarify the dominant 
energy transfer pathways, which sensitively depend on the 
anharmonic couplings among vibrational modes. 
A ratio of the anharmonic coupling to the frequency mismatch
is employed to predict and interpret the dominant energy flow pathways.

\end{abstract}

\end{frontmatter}

\section{Introduction}

The dynamics of proteins involves high frequency vibrational modes 
that behave quantum mechanically even at room temperature
($T \simeq 300$ K corresponds to $\simeq200$ cm$^{-1}$).
One such class of vibrations is the amide I mode, 
centered around 1650 cm$^{-1}$ and prominent
in IR experiments due to its large oscillator strength. 
The amide I modes have been intensively studied by 2D-IR 
spectroscopy, which has provided new insights into their anharmonic 
couplings to the other vibrational modes and their intrinsic mode 
anharmonicities \cite{ZH01,WH06,Tokmakoff06}.
Theoretical studies have also been stimulated to characterize
the details of the vibrational states \cite{GCG02,Mukamelgroup,BS06,KB07}, 
and to probe the origin of the inhomogeneous lineshape 
broadening of the amide I mode \cite{Mukamelgroup,HHC05,Skinnergroup,Stock06}. 

In this study, we focus on the vibrational energy relaxation (VER) 
process of the amide I mode. 
Recent pump-probe experiments \cite{HLH98,ZAH01} have found that the VER of the 
amide I mode occurs on a a sub-picosecond timescale after its excitation. 
Nguyen and Stock \cite{NS03} have studied this phenomenon using molecular 
dynamics and instantaneous normal mode analysis. 
Their results were found to be in qualitative agreement with the experimental 
results. Nevertheless a more precise quantum mechanical description of the VER process 
is desirable.
Two of the authors \cite{FZS06} have recently proposed a 
time-dependent perturbation theory 
to describe VER, 
and applied the formula to a small peptide-like molecule, N-methylacetamide (NMA), 
in heavy water using an empirical force field.
They also observed VER on a time scale ($\sim 0.5$ ps) similar to that obtained 
experimentally, and further proposed a mechanism for the VER process.

The present study explores the VER processes of 
NMA {\it more accurately} by directly solving the Schr\"odinger equation 
for molecular vibrations on {\it ab initio} potential energy surface (PES). 
The NMA molecule \textit{in vacuum} is 
examined, excluding water molecules included in the previous studies \cite{NS03,FZS06}.
Quantum dynamics calculations are carried out by the vibrational 
configuration-interaction (VCI) method \cite{bowman1,gerber,MULTIMODE,bowman2}.
24 vibrational modes (out of 30) are explicitly treated in the 
dynamics, excluding the 6 lowest-lying modes including the rotational 
motions of the two methyl groups. 
We note that while the multiconfiguration time-dependent Hartree (MCTDH) method 
\cite{BJWM00} provides a more flexible framework the VCI method 
is sufficient for the present aim to investigate the vibrational 
motion of a semi-rigid molecule.


\section{Methods}

\subsection{Quartic force field}
\label{sec:3MR}
The PES is approximated using a fourth-order Taylor series 
expansion around the equilibrium geometry, which is called 
quartic force field (QFF), as
\begin{equation}
\tilde{V}(\{Q_i\})
=
\frac{1}{2}\sum_k \omega_k^2 Q_k^2 
+ \frac{1}{3!}\sum_{k,l,m} t_{klm} Q_kQ_lQ_m  
+ \frac{1}{4!}\sum_{k,l,m,n}u_{klmn} Q_k Q_l Q_m Q_n,
\end{equation}
where $Q_k$ and $\omega_k$ denote the $k$th normal coordinate and the 
associated harmonic frequency. The coefficients, $t_{klm}$ and $u_{klmn}$,
are the third- and fourth-order derivatives of the PES. 
The above QFF can be 
recast in the form of 
the $n$-mode coupling representation ($n$MR) \cite{YTHG01,YHTSG04} as
\begin{eqnarray}
\tilde{V}(\{Q_i\})
&\simeq&
V(\{Q_i\})
=
V^{1MR}+V^{2MR}+V^{3MR},
\\
V^{1MR}(\{Q_i\})
&=&
\sum_k 
\left(
\frac{1}{2} \omega_k^2 Q_k^2
+\frac{1}{3!} t_{kkk} Q_k^3
+\frac{1}{4!} u_{kkkk} Q_k^4
\right),
\\
V^{2MR}(\{Q_i\})
&=&
\sum_{k,l} 
\left( \frac{1}{3!} t_{kkl} Q_k^2 Q_l + \frac{1}{4!} 
u_{kkll} Q_k^2 Q_l^2 + \frac{1}{4!} u_{kkkl} Q_k^3 Q_l
\right),
\\
V^{3MR}(\{Q_i\})
&=&\sum_{k,l,m} 
\left(
\frac{1}{3!} t_{klm} Q_k Q_l Q_m + \frac{1}{4!}u_{kklm} Q_k^2 Q_l Q_m 
\right). 
\end{eqnarray}
It is known that 3MR-QFF is enough to characterize the anharmonicity
of molecules studied \cite{YTHG01,YHTSG04}, we thus 
have neglected the fourth-order terms including
$u_{klmn}$ with $k \neq l \neq m \neq n$.

In addition to the above 3MR-QFF, we study two types of approximate QFF:  
(1) 2MR-QFF that neglects $V^{3MR}$ \cite{GCG02}
and 
(2) partial-QFF that takes into account only the terms
associated with the amide I mode, that is, $t_{klm} = 0$ and 
$u_{kklm}=0$ if the subscripts do not include the amide I mode.
Note that the previous perturbation calculation 
\cite{FZS06} also employed partial-QFF to calculate the 
reduced density matrix.
By adding the normal mode kinetic energy $K=\sum_k P_k^2/2$,
we have the full approximate vibrational 
Hamiltonian $H=K+V(\{ Q_i \})$ for the system.


\subsection{Vibrational CI method}


The vibrational self-consistent field (VSCF) calculation is first 
carried out for the vibrational ground state to obtain the basis 
functions for the VCI calculations. 
The VSCF wavefunction is expressed as a direct product of one-mode 
functions or {\it modals} as
\begin{eqnarray}
 \Psi_{\mathbf{n}}^{\mathrm{VSCF}} = \prod_{i=1}^{f} \phi_{n_i}^{(i)}(Q_i),
\end{eqnarray}
where {\bf n} and $f$ denote the vibrational quantum numbers and 
the degrees of freedom, respectively. The modals are determined by 
\begin{eqnarray}
 \left[ -\frac{1}{2}\frac{\partial^2}{\partial Q_i^2} + 
 \langle \prod_{j \neq i} \phi_{n_j}^{(j)} |
 V | \prod_{j \neq i} \phi_{n_j}^{(j)} \rangle \right] 
 \phi_{n_i}^{(i)} =
 \epsilon_{n_i}^{(i)} \phi_{n_i}^{(i)}. 
\end{eqnarray}
This VSCF equation is solved for the vibrational ground state ({\bf n=0}), 
and the $virtual$ modals constitute the VSCF configurations to be used 
in the VCI calculations. The VCI wavefunction is expressed as a 
linear combination of VSCF configurations as
\begin{eqnarray}
 \Psi_{\mathbf{n}}^{\mathrm{VCI}} = \sum_{\mathbf{m}} C_{\mathbf{mn}} 
 \Psi_{\mathbf{m}}^{\mathrm{VSCF}}.
\end{eqnarray}
The VCI wavefunction and energy levels are obtained by diagonalization of the 
VCI matrix
\begin{eqnarray}
 H_{\mathbf{mn}} = \langle \Psi_{\mathbf{m}}^{\mathrm{VSCF}} | H 
 | \Psi_{\mathbf{n}}^{\mathrm{VSCF}} \rangle.
\label{eq:VCImatrix}
\end{eqnarray}

In this study, the modals were expanded in terms of the harmonic 
oscillator (HO) wavefunction.  The number of HO wavefunctions employed 
were 11, 9, 7, and 5 for \{$\phi^{(7)},\phi^{(8)}$\}, \{$\phi^{(9)}$-
$\phi^{(12)}$\}, \{$\phi^{(13)}$-$\phi^{(23)}$\}, and 
\{$\phi^{(24)}$-$\phi^{(30)}$\}, respectively. 
The mode index is 
labeled in the increasing order of the frequency. 
See Table \ref{tab:freq2}.
The 6 lowest-lying modes were kept frozen. The VSCF 
configurations were constructed by allowing the excitation up to 10 
quantum numbers, and selected in the increasing order of 
the energy until the upper limit of the VCI space, denoted $N_{CI}$, 
was achieved.  The VSCF/VCI calculations were carried out using 
the S{\footnotesize INDO} code \cite{sindo} for non-rotating molecules.


\begin{table}[htbp]
\caption{
Calculated harmonic (H.O) and anharmonic (VCI and PT2) frequencies of 
NMA based on the B3LYP/6-31G+(d) level of theory. The previous cc-VSCF and VCI 
results and the experimental results are also listed for comparison. 
Units in cm$^{-1}$.
}
\begin{center}
\begin{tabular}{c c c c c c c}
\\ 
\hline \hline
Mode  & H.O.$^a$ & VCI$^b$ & PT2$^b$ & cc-VSCF$^c$  & VCI$^d$ & Exp.$^e$ 
 \\ \hline \hline           
7  & 623  & 619  & 614  & 636    & 633    & 619 \\
8  & 630  & 625  & 613  & 637    & 685    & 658 \\
9  & 879  & 869  & 861  & 891    & 886    & 857 \\
10 & 1001 & 983  & 982  & 1022   & 1005   & 980 \\
11 & 1066 & 1046 & 1042 & 1083   & 1061   & 1037 \\
12 & 1107 & 1093 & 1080 & 1119   & 1099   & 1089 \\
13 & 1163 & 1141 & 1126 & 1184   & 1167   & ---  \\
14 & 1199 & 1186 & 1146 & 1214   & 1199   & 1168 \\
15 & 1292 & 1272 & 1256 & 1283   & 1253   & 1266 \\
16 & 1421 & 1397 & 1387 & 1421   & 1388   & 1370 \\
17 & 1473 & 1448 & 1432 & 1468   & 1442   & 1419 \\
18 & 1495 & 1451 & 1481 & 1515   & 1467   & 1432 \\
19 & 1500 & 1453 & 1459  & 1557  & 1483  & 1432 \\
20 & 1507 & 1469 & 1476  & 1541  & 1481  & 1446 \\
21 & 1529 & 1495 & 1481  & 1566  & 1505  & 1472 \\
22 & 1560 & 1537 & 1505  & 1547  & 1519  & 1511 \\
23 & 1751 & 1725 & 1725  & 1751  & 1727  & 1707 \\
24 & 3059 & 3017 & 2906  & 2939  & 2940  & 2915 \\
25 & 3060 & 2996 & 2940  & 2940  & 2983  & 2958 \\
26 & 3117 & 3077 & 2963  & 2979  & 2995  & 2973 \\
27 & 3132 & 3093 & 2977  & 3014  & 3043  & 3008 \\
28 & 3136 & 3100 & 2990  & 2985  & 3025  & 2973 \\
29 & 3149 & 3077 & 2990  & 2993  & 3056  & 3008 \\
30 & 3643 & 3479 & 3466  & 3523  & 3544  & 3498 \\
\hline \hline 
\end{tabular}
\end{center}
\label{tab:freq2}
\vspace{0.3cm}
$^a$ Harmonic frequencies at the B3LYP/6-31+G(d) level. \\
$^b$ Based on 3MR-PES at the B3LYP/6-31+G(d) level. \\
$^c$ Reference \cite{GCG02}. Based on 2MR-PES at the MP2/DZP level. \\
$^d$ Reference \cite{KB07}. Based on partial-3MR-PES at the MP2/aug-cc-pVTZ level. \\
$^e$ Reference \cite{ATT84}.

\end{table}

\subsection{Quantum dynamics}

Once the eigenvalues (\{$E_\mathbf{n}$\}) and the 
eigenfunctions (\{$\Psi_\mathbf{n}^\mathrm{VCI}$\}) are obtained, 
it is straighforward to carry out an approximate quantum dynamics simulation
\begin{eqnarray}
|\Psi(t) \rangle
=
\sum_\mathbf{n} \langle \Psi_\mathbf{n}^\mathrm{VCI} | \Psi(0) \rangle 
e^{-iE_\mathbf{n}t/\hbar} 
| \Psi_\mathbf{n}^\mathrm{VCI} \rangle.
\end{eqnarray}
If we take the initial state to be the VSCF configuration,
$\Psi(0) = \Psi_\mathbf{i}^\mathrm{VSCF}$, then
the overlap integral between $\Psi(t)$ and 
$\Psi_\mathbf{j}^\mathrm{VSCF}$ is calculated as
 \begin{eqnarray}
O_\mathbf{j}(t)
\equiv
\langle \Psi_\mathbf{j}^\mathrm{VSCF} |\Psi(t) \rangle
=
\sum_\mathbf{n} 
C_\mathbf{jn} C_\mathbf{ni}
e^{-iE_\mathbf{n} t/\hbar}. 
 \end{eqnarray}
The absolute square of $O_\mathbf{j}(t)$ gives the probability of {\bf j} 
at time $t$, $P_\mathbf{j}(t) = |O_\mathbf{j}(t)|^2$. 
Dynamics calculations were carried out with the initial 
VSCF configurations corresponding to the fundamental and 
first overtone of the amide I mode. 
We note that we also examined different initial states which are 
the superposition of states near the amide I mode fundamental or 
overtone, but the results do not severely depend on how to 
prepare the initial states \cite{FYHS07}.

\section{Results}

\subsection{Anharmonic frequency calculations of NMA: Accuracy of the PES}
\label{sec:accuracy}

Here we shall examine the accuracy of the PES derived from B3LYP/6-31+G(d) 
employing anharmonic frequency calculations 
because the vibrational quantum dynamics sensitively depends on  
anharmonicity of a system. 
We compare our result with the previous 
theoretical values obtained by cc-VSCF method \cite{GCG02} 
and very recently by VCI method \cite{KB07} using MULTIMODE \cite{MULTIMODE} 
as well as with the experimental values \cite{ATT84}.
We have also computed the fundamental frequencies of NMA by the second-order 
perturbation theory (PT2) \cite{Barone,GWGH90} using Gaussian03 \cite{Gaussian03}.
Table \ref{tab:freq2} shows the resulting harmonic (H.O.) and anharmonic 
frequencies (VCI and PT2) based on the B3LYP/6-31+G(d) level of theory, 
together with the previous cc-VSCF results based on the MP2/DZP \cite{GCG02} 
and the VCI results based on the MP2/aug-cc-pVTZ \cite{KB07}.
The experimental results are taken from infrared spectra of NMA 
in nitrogen matrix \cite{ATT84}. 
Although the experiment was not in the gas phase, 
we expect that the nitrogen matrix should affect minimally for  
vibrational frequencies, and we regard these correct 
values.
As is well known, the harmonic frequencies overestimate the 
correct frequencies but the anharmonic frequencies are 
closer to the latter. A close inspection shows that the 
PT2 result is the most accurate but marginally 
and the other three methods are almost comparable except for 
six CH stretching modes ($\sim 3000$ cm$^{-1}$), where 
our result seems to be crudest.
However, we are interested in the amide I mode, whose frequency 
is around 1700 cm$^{-1}$, and the VER pathways from the amide I mode 
are usually toward lower frequency modes, 
in this paper we accept this level of accuracy. 

\subsection{Vibrational quantum dynamics of NMA-D: Fermi resonance}

\begin{figure}[htbp]
\hfill
\begin{center}
\includegraphics[scale=1.0]{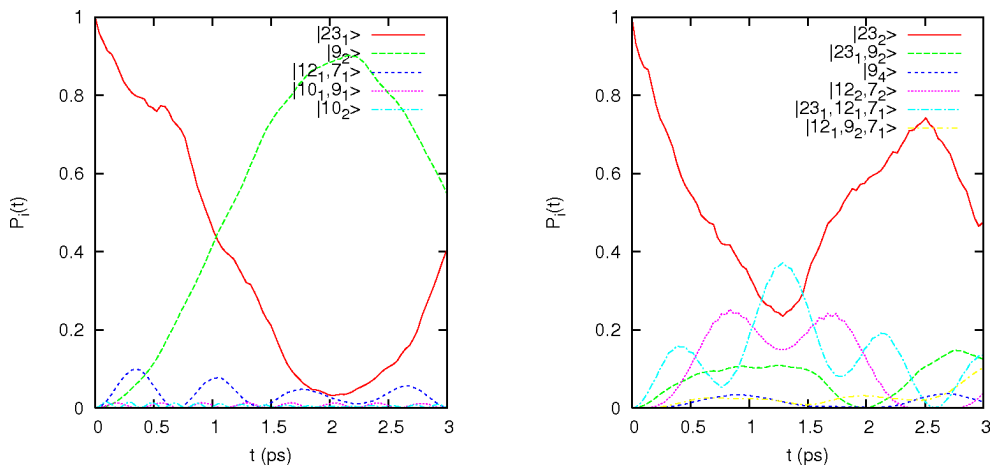}
\end{center}
\caption{
Quantum population dynamics of the vibrational excited 
states, $v=1$ (left) and  $v=2$ (right) of the amide I mode, 
on 3MR-QFF derived from B3LYP/6-31G+(d) method. 
$N_{CI}$ is set to 6000. 
The dominant VER pathways are also 
summarized in Table \ref{tab:pathway}. 
}
\label{fig:label}
\end{figure}

To compare with the previous dynamics calculation using a force field \cite{FZS06} and 
time-resolved experiments \cite{HLH98,ZAH01}, 
we deuterated NMA into NMA-D (CH$_3$-ND-CO-CH$_3$).
We show the VER dynamics of NMA-D calculated by VCI method in Fig.~\ref{fig:label}.
We consider two initial states: $v=1$ (fundamental) and $v=2$ (overtone) 
excitation of the amide I mode in the VSCF base.
In both cases,
the initial ``decay'' appears to occur on a sub picosecond timescale,
which is in accord with the previous studies \cite{ZAH01,FZS06}.
However, the long-time dynamics (not shown) are 
quasi-periodic with resonant energy ``transfer'' between the amide I 
mode and the other modes.  In order to induce irreversible decay,
it is essential to include more bath modes in the form of solvent molecules 
so that energy can dissipate \cite{FZS06}.  A vibrationally excited 
molecule with sufficient quanta can decay irreversibly even in a vacuum \cite{SWW96}.  
The present result indicates that 
the intramolecular energy ``transfer'' process is 
responsible for the initial fast ``decay'' ($\le  1$ ps).

\begin{figure} 
\hfill
\begin{center}
\includegraphics[scale=0.8]{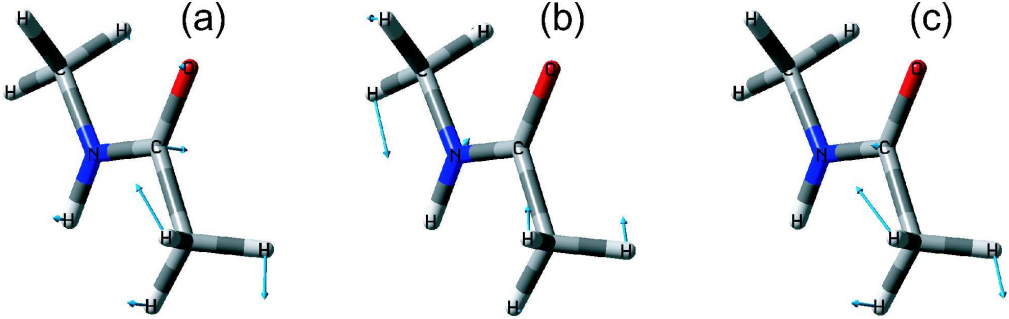}
\end{center}
\caption{
Normal mode vectors of the resonant bath modes 
for the amide I mode in deuterated NMA. 
(a) 7th (619 cm$^{-1}$), 
(b) 9th (869 cm$^{-1}$), and 
(c) 12th (1066 cm$^{-1}$) mode.
}
\label{fig:resmode}
\end{figure}

The dominant VER pathways are described 
in Table \ref{tab:pathway}, and the resonant bath modes 
are depicted in Fig.~\ref{fig:resmode}.
We interpret these dominant pathways 
using the following Fermi resonance parameter \cite{Cremeens06}
\begin{equation}
\eta \equiv
\left| 
\frac{\langle i| \Delta V|f \rangle}{\Delta E}
\right|
\end{equation}
where $|i \rangle$ and $|f \rangle$ are 
the initial and final {\it harmonic} states, 
$\Delta V = V-\sum_k \omega_k^2 Q_k^2 /2$ 
is the anharmonic potential energy, 
and $\Delta E$ is the energy difference between 
$|i \rangle$ and $|f \rangle$.
For example, $\eta$ for a transition 
between $|i \rangle= |S_1 \rangle$ (the  
system mode is singly excited) and $|f \rangle=|k_1 l_1 \rangle$ 
(two bath modes are singly excited) is evaluated as 
\begin{equation}
\eta 
= 
\left| 
\frac{\langle S_1 | 2 t_{Skk} Q_S Q_k Q_l | k_1 l_1 \rangle}
{6\hbar (\omega_S -\omega_k-\omega_l)}
\right| 
=
\left| 
\frac{t_{Skl}}
{3 \hbar(\omega_S - \omega_k-\omega_l)}
\right|
\sqrt{\frac{\hbar}{2 \omega_S}}
\sqrt{\frac{\hbar}{2 \omega_k}}
\sqrt{\frac{\hbar}{2 \omega_l}}.
\end{equation}
Note that the Fermi resonance parameter was a key ingredient of 
the time-dependent perturbation theory 
in describing the reduced density matrix for 
the amide I mode \cite{FZS06}. Similar analysis 
has been applied to protein dynamics of myoglobin 
in terms of classical mechanics \cite{MMK00}.

\begin{table} 
\hfill
\caption{
The dominant VER pathways in NMA-D and the 
corresponding Fermi resonance parameters 
when we excite the fundamental of the amide I mode.
$|m_n \rangle$ denotes the $n$th excited states
of the $m$th mode, and all other modes are on the 
vibrational ground state. The 23rd mode is the amide I mode 
in this paper. 
The result is derived from the B3LYP/6-31G+(d)/3MR-PES.
 }
\begin{center}
\begin{tabular}{c c}
\\ 
\hline \hline
VER pathway & Fermi resonance parameter  
\\ \hline \hline           
$|23_1 \rangle \rightarrow |9_2 \rangle$ & 0.082 \\ 
$|23_1 \rangle \rightarrow |12_1 7_1 \rangle$ & 0.024 \\ 
$|23_1 \rangle \rightarrow |10_2 \rangle$ & 0.014 \\ 
$|23_1 \rangle \rightarrow |10_1 9_1 \rangle$ & 0.013 \\ 
\hline \hline 
\end{tabular}
\end{center}
\label{tab:pathway}
\end{table}

\begin{figure}[htbp]
\hfill
\begin{center}
\includegraphics[scale=0.9]{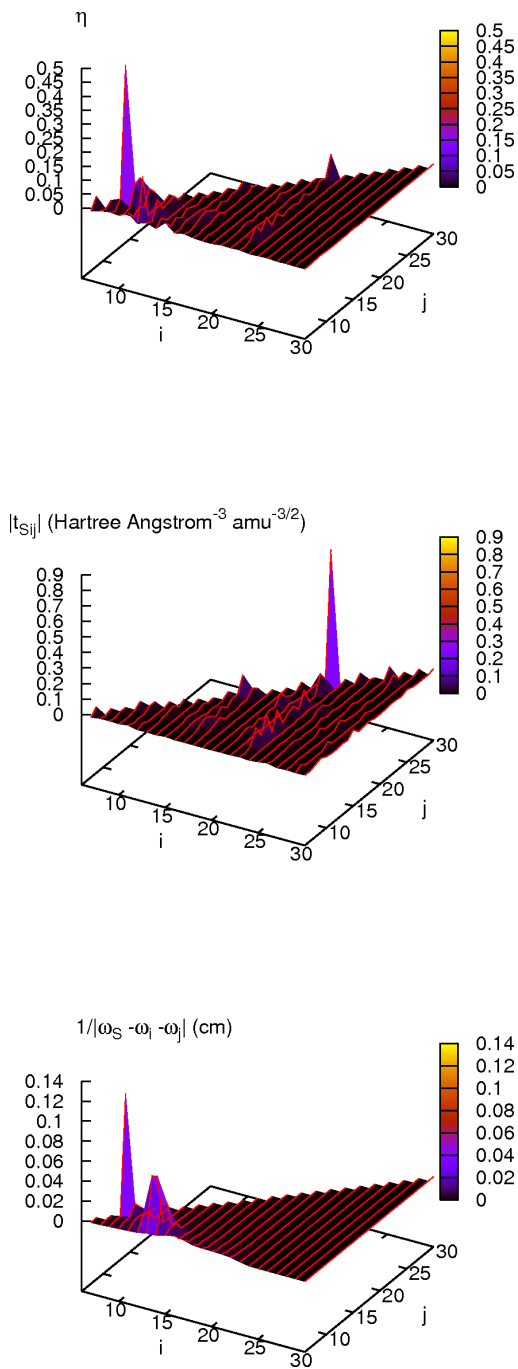}
\end{center}
\caption{
Top: 
The Fermi resonance parameter 
as a function of 
the mode index $(i,j)$.
Middle: The 3rd order coupling strength $|t_{Sij}|$ 
as a function of the mode index $(i,j)$.
Bottom: The inverse of the frequency mismatch 
$1/|\omega_S-\omega_i-\omega_j|$ 
as a function of the mode index $(i,j)$.
}
\label{fig:FRP}
\end{figure}

Table \ref{tab:pathway} also shows 
that the dominant VER pathways are well characterized by 
the above parameters for the $v=1$ excitation.  
Figure \ref{fig:FRP} shows that
other minor pathways take on values of $\eta$ that are 
smaller than $\sim$ 0.01 
and that the frequency matching condition is 
important to predict the Fermi resonance parameter.
It is interesting to note that the dominant pathways 
(Fig.~\ref{fig:resmode})
are similar to those suggested in \cite{HLH98}. 
In the case of $v=2$ excitation (Fig.~\ref{fig:label}, right), 
it is difficult to assign such a parameter 
because the relaxation processes can accompany
higher order processes, but 
the dominant pathways such as 
$|23_2 \rangle \rightarrow |23_1, 12_1, 9_1\rangle$ 
and 
$|23_2 \rangle \rightarrow |9_4\rangle$ are 
the frequency matching ones. 
We conclude that the Fermi resonance parameter or 
frequency matching condition is very important 
to characterize the VER pathways, and its consequence is 
discussed below.

\subsection{Discussions}

\begin{figure} 
\hfill
\begin{center}
\includegraphics[scale=1.0]{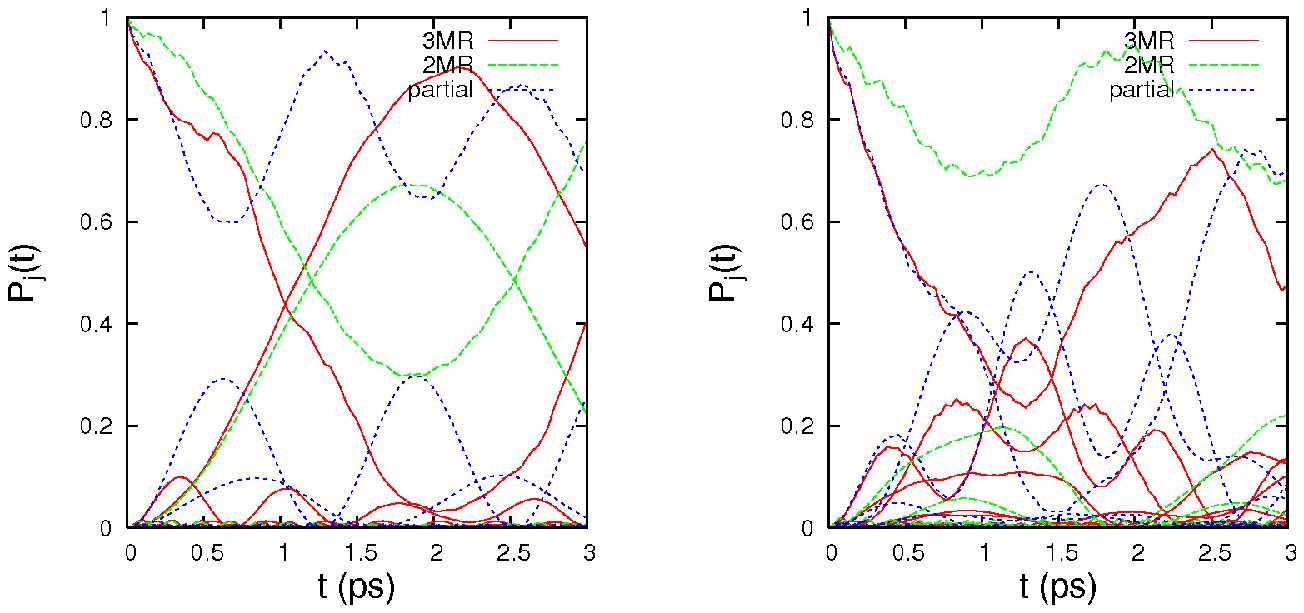}
\end{center}
\caption{
Quantum population dynamics of the vibrational excited 
states, $v=1$ (left) and  $v=2$ (right) of the amide I mode,  
on 3MR-QFF, 2MR-QFF, and partial-QFF, derived from 
B3LYP/6-31G+(d) method. 
$N_{CI}$ is set to 6000. 
}
\label{fig:B3LYPdynashort}
\end{figure}

From these observations, 
we conclude that a relatively small number of resonant bath modes plays 
an essential role in the quantum dynamics. 
Simplification of the PES 
or a reduction in the number of modes, including the resonant bath modes,
can result in an inaccurate estimate of the dynamics.
This is illustrated in Fig.~\ref{fig:B3LYPdynashort}:
the 3MR-QFF and partial-QFF results agree rather well 
up to 0.2 (0.5) ps for $v=1$ ($v=2$), 
while 2MR-QFF results deviate from the other two 
at the initial stage. 
This is because the 2MR representation misses resonances 
mediated through  the three mode interactions. 
It is interesting that 
the vibrational frequencies can be calculated accurately using 2MR-QFF; the 
fundamental (first overtone) of the amide I mode is obtained as 1713 (3414), and 
1716 (3421) cm$^{-1}$ by the 2MR- and 3MR-QFF, respectively 
(the difference is 0.2 \%). 
However, the VER processes
sensitively depend on the 3MR terms. 


We showed that our PES is accurate enough (Sec.~\ref{sec:accuracy}), 
which is in accord with experiment \cite{ATT84} 
and comparable to other theoretical methods \cite{Gaussian03,GCG02,KB07}.
However, we note that the VER dynamics 
are much more sensitive to accuracy
than the anharmonic frequency calculations. 
For example, Gerber's group did not report any strong resonant 
interaction between the amide I mode and other modes 
using MP2/DZP level of theory \cite{GCG02}.
This is likely due to the level of theory for the PES (MP2/DZP), 
but another possibility is due to the 2MR-PES in their calculations.
As we showed above, 3MR-PES and 2MR-PES can provide similar anharmonic 
frequencies, but some resonance conditions should be missed in the latter, 
affecting the dynamics calculations. 
Importantly, 2D-IR spectroscopy can directly detect anharmonic 
coupling in a molecule \cite{ZH01,WH06,Tokmakoff06}, and 
such data may be utilized to 
select an {\it appropriate} level of methods for a particular molecule.

\section{Concluding remarks}

Using the VCI method, we investigated the quantum dynamics of 
deuterated N-methylacetamide in vacuum.
We demonstrated the applicability of 
the method and were able to gain insight into the 
fundamental nature of VER in the molecule, 
relevant to the interpretation of IR and 2D-IR spectroscopy used as 
a probe of protein dynamics.
The accuracy of the PES employed (B3LYP/6-31+G(d)) 
was checked by the anharmonic frequency calculations
and by comparing with experiment and other theoretical methods.
We clarified the energy flow pathways from the 
$v=1$ and 2 excitations of the amide I mode, and 
interpreted our results using Fermi resonance 
parameters, which represent the effective coupling strength 
betweem vibrational modes.
This approach will be extended to condensed phase systems 
by invoking QM/MM methods \cite{LSCLS06,Hirata05} 
or multiresolution methods \cite{BS06,rauhut,YHH06} 
to deal with dephasing problems and to simulate 2D-IR signals \cite{Tanimura06}.



{\bf Acknowledgments}

We are grateful to E.~Geva, G.~Stock, D.M.~Leitner,
J.M.~Bowman, and Y.~Zhang for useful discussions.
We also thank the National Science 
Foundation (CHE-0316551) and Boston University's Center for Computer Science 
for generous support to our research.
JES is grateful to Montana State University for generous support and 
hospitality.



\end{document}